\documentclass[preprint,jgrga]{agutex}
\usepackage{amssymb}

\usepackage[dvips]{graphicx}

\setkeys{Gin}{draft=false}

\authorrunninghead{VAN SUMMEREN ET AL.}
\titlerunninghead{DYNAMO LIFETIMES IN ROCKY EXOPLANETS}

\authoraddr{
Clinton P. Conrad,
Department of Geology and Geophysics, 
University of Hawaii, 
1680 East-West Road, 
Honolulu, 
HI 96822, 
U.S.A.
} 

\authoraddr{
Eric Gaidos,
Department of Geology and Geophysics, 
University of Hawaii, 
1680 East-West Road, 
Honolulu, 
HI 96822, 
U.S.A.
}

\authoraddr{
Joost van Summeren,
Department of Geology and Geophysics, 
University of Hawaii, 
1680 East-West Road, 
Honolulu, 
HI 96822, 
U.S.A.
(summeren@hawaii.edu)
}

\begin{document}
\title{Magnetodynamo Lifetimes for Rocky, Earth-Mass Exoplanets with
  Contrasting Mantle Convection Regimes}

\authors{
Joost van Summeren,\altaffilmark{1}
Eric Gaidos,\altaffilmark{1}
Clinton P. Conrad,\altaffilmark{1} 
}

\altaffiltext{1}{Department of Geology and Geophysics, University of Hawaii, Honolulu, Hawaii, U.S.A.}

%% ------------------------------------------------------------------------ %%
\begin{abstract}
We used a thermal model of an iron core to calculate magnetodynamo evolution in Earth-mass rocky planets to determine the sensitivity of dynamo lifetime and intensity 
to planets with different mantle tectonic regimes, surface temperatures, and core properties. 
The heat flow at the core-mantle boundary (CMB) is derived from numerical models of mantle convection with a viscous/pseudo-plastic rheology that captures the phenomenology of 
plate-like tectonics.
Our thermal evolution models predict a long-lived ($\sim$8 Gyr) field for Earth and similar dynamo evolution for Earth-mass exoplanets with plate tectonics.
Both elevated surface temperature and pressure-dependent mantle viscosity reduce the CMB heat flow but produce only slightly longer-lived dynamos ($\sim$8--9.5 Gyr).
Single-plate (``stagnant lid'') planets with relatively low CMB heat flow produce long-lived ($\sim$10.5 Gyr) dynamos.
These weaker dynamos can cease for several billions of years and subsequently reactivate due to the additional entropy production associated with inner core growth, a possible explanation for the absence of a magnetic field on present-day Venus.
We also show that dynamo operation is sensitive to the initial temperature, size, and solidus of a planet's core. 
These dependencies would severely challenge any attempt to distinguish exoplanets with plate tectonics and stagnant lids based on the presence or absence of a magnetic field.
\end{abstract}

%% ------------------------------------------------------------------------ %%
\begin{article}

%%%%%%%%%%%%%%%%%%%%%%%%%%%%%%%%%%%%%%%%
\section{Introduction}

Many planets in the Solar System have or had magnetic fields produced by internal dynamos and this is presumably also true among planets orbiting other stars.
Magnetic fields of exoplanets have yet to be unambiguously detected, but corotating regions of elevated chromospheric emission on the host stars of some close-in giant planets can be explained by magnetic interaction \citep{Shkolnik2005,Shkolnik2008,Fares2012} and one explanation for the inflated radii of many giant planets on close-in orbits is heating by ohmic dissipation during circulation of a partially ionized upper atmosphere through the planet's magnetic field \citep{Batygin2010}.
Past or present operation of magnetodynamos has been linked to the interior structure and thermal evolution of the rocky planets of the inner Solar System \citep{Breuer_etal2010,Stevenson2010}  and the detection of magnetic fields might, in principle, provide insight into the interiors of exoplanets.

In differentiated silicate-metal planets like the Earth, dynamos are produced by convection within the liquid or partially liquid iron core if the magnetic Reynolds number $Re_m > 40$ \citep{Stevenson2010}.  
Dynamo operation is associated with ohmic dissipation, which requires entropy production in excess of that from conduction along the core's adiabat.
Entropy production can be due to
 cooling of the core in excess of conduction, release of latent heat during solidification of an inner core, or formation of buoyant fluid generated by the exclusion of lighter elements such as sulfur when iron crystallizes.  
All three sources require removal of heat from the
core and thus core cooling is considered a requirement for dynamo
operation.  The decay of the long-lived radioactive isotope $^{40}$K is
another possible energy source
\citep{Lee2003,Murthy_etal2003,Nimmo2004} although  little such potassium may partition into a core \citep{Corgne2007}.  
Planetary rotation rate is thought to have little influence on dynamo strength although it may control 
the relative strength of the dipole and multipolar components \citep{Zuluaga2012}.

In an Earth-like planet, the core is surrounded by a silicate mantle
and heat must be carried from the core-mantle boundary (CMB) by
subsolidus convection.  
The viscosity of a silicate mantle is strongly temperature-dependent and most likely the mantle contains long-lived radionuclides that provide heat and sustain high mantle temperatures over billions of years.  
The ability of the mantle to carry heat from  the core depends also on the style of mantle convection, distinguished by the degree to which the brittle lithosphere participates in convection, e.g. plate tectonics or stagnant lid modes.
Thus, magnetodynamo operation is tied to the tectonics and thermal evolution
of the planet's mantle.  
The lack of a dynamo in Venus has been ascribed to the apparent absence of plate tectonics \citep{Buffett_etal1996,Nimmo2002a}.

Previous theoretical studies have predicted the evolution of dynamos in rocky exoplanets \citep{Gaidos_etal2010,Driscoll2011,Tachinami2011}.
\citet{Gaidos_etal2010} examined the effect of surface temperatures and plate tectonics. 
They found that higher surface temperatures can enhance the dynamo due to the lower viscosity of a hotter mantle and that dynamos are weaker but still operate on planets with stagnant lids.  
To explain the absence of a present-day dynamo on Venus it was necessary to invoke an 
order-of-magnitude enhancement in mantle viscosity.
They used parametric models to describe mantle convection and heat transport across the CMB.  
However, accounting for the complexity of subsolidus convection in a planetary mantle with a strongly temperature-dependent viscosity requires numerical simulations. 

Furthermore, numerical models can adequately treat the effects of a non-uniform surface temperature.
Rocky planets very close to their parent star will be tidally locked,
most likely with a 1:1 spin-orbit ratio, and may lack an atmosphere
due to erosion by the stellar wind \citep{Khodachenko2007}.  In this
case, temperatures will vary markedly between the irradiated and dark
sides, and possible patterns of mantle convection include a degree-one
(asymmetric) mode \citep{vanSummeren_etal2011}.  
This could enhance heat flow inhomogeneity across the CMB and suppress a dynamo 
if peak-to-peak variations are of the same order as mean values \citep{OlsonChristensen2002}.

Finally, the rheology of planetary mantles is expected to be pressure-dependent as well as temperature-dependent, although the pressure-dependence for super-Earths is a topic of controversy.
If viscosity increases with pressure \citep{Stamenkovic_etal2011}, this will suppress convective vigor in the deep mantles of massive planets \citep{Wagner_etal2011,Tackley2012}, which in turn will reduce heat flow across the CMB.
\citet{Tachinami2011} predict that, above a critical planetary mass, enhanced viscosity terminates dynamo operation.
On the other hand, \citet{Karato2011} argues for a {\it decrease} in viscosity with pressure by 2 to 3 orders of magnitude due to changes in the diffusion mechanism (from vacancy to interstitial) and a series of chemical and electronic phase transitions over 0.1--1 TPa.  
If this is correct, the deep mantles of super-Earths will be less viscous and layered convection may occur.

To more thoroughly investigate the conditions for the appearance of dynamos in Earth-like exoplanets, we investigated the influence of differences in convective regime, surface temperature, pressure-dependence of mantle viscosity, and a range of core properties (radius, temperature, thermal conductivity, and solidus) on the lifetimes and intensities of magnetodynamos in planets with the same mass and interior structure as Earth.
We show how dynamos can be longer-lived for planets with immobile surfaces compared to those with plate tectonics.
We then determined the sensitivity of dynamo behavior to the planetary and physical parameters of our simulations, and assessed how informative dynamo-generated fields may be for probing the interior dynamics of exoplanets.

%%%%%%%%%%%%%%%%%%%%%%%%%%%%%%%%%%%%%%%%
\section{Methods}
\label{Sect-Methods}

\subsection{General modeling approach}

To explore parameter space, we calculated a series of dynamo evolution calculations for different mantle tectonic regimes, surface temperatures, and core properties.
A corresponding number of forward calculations for mantle evolution over billions of years is computationally impractical. 
Therefore, we adopted the following approach: 
(i) We calculated the steady state CMB heat flow, $\overline{Q_{CMB}}$, from numerical models of mantle convection with an imposed constant CMB temperature, $T_{CMB}$, and rate of mantle internal heating, $H_m$.
(ii) We repeated this procedure for different CMB temperatures to derive parameterizations of $\overline{Q_{CMB}}$ vs. $T_{CMB}$ for each convection scenario. 
(iii) We employed these parameterizations in an analytical model for the thermal evolution of an iron core, which we solved numerically. 
In these models, we rescaled the $\overline{Q_{CMB}}$ values to account for time-dependent mantle internal heating rates, consistent with boundary layer theory. 
(iv) We calculated dynamo intensities using a scaling law with core entropy terms. 

\subsection{Thermal model of the core}
\label{Sect-coremodel}

We consider liquid/solid Fe cores with the same total radius of that of Earth, except in Section \ref{Sect-res-life} where we vary the core radius. 
To estimate the intensity of a magnetic dipole field at the CMB ($B_{CMB}$) we adopt the following scaling
law derived from numerical dynamo simulations
\citep{ChristensenAubert2006,OlsonChristensen2006,Aubert_etal2009}:
\begin{equation}
\label{Eq-scalinglaw}
B_{CMB} = \alpha \sqrt{\mu_0 \bar{\rho}}
\left[ \frac{\Phi\bar{T}}{M_c}(r_{CMB}-r_{i}) \right]^{1/3},
\end{equation}
with $\Phi$ the entropy production rate associated with ohmic dissipation due to electrically charged currents in a conducting fluid, 
$\mu_0$ the magnetic diffusivity, $\bar{\rho}$ the average core density, $\bar{T}$ the temperature at which ohmic dissipation occurs \citep{Roberts_etal2003}, $M_c$ the core mass, and $r_{CMB}$ and $r_{i}$ the radii of the CMB and inner-outer core boundary, respectively.  
Parameters and values are listed in Table \ref{tab-notation}.  
The dimensionless coefficient $\alpha=c_1 \cdot b_{dip}=0.165$ combines a prefactor used to extrapolate dynamo modeling results from comparatively small Rayleigh numbers and large Ekman and magnetic Prandtl numbers to planetary conditions ($c_1$=1.65, \citet{Aubert_etal2009}) and a fitting coefficient that relates the mean field strength inside the core to the dipole field at the CMB ($b_{dip}\approx10$, \citet{Aubert_etal2009}). 
The adopted scaling law successfully predicts the magnetic moments of the Solar System planets and Ganymede \citep{OlsonChristensen2006}.
For pure dipoles, the average field intensities at the planetary surface and CMB are related through $B_{srfc}\approx(r_{CMB}/r_{srfc})^3 B_{CMB}$.  

We calculated the entropy production rate associated with ohmic dissipation, $\Phi$, from the time-dependent
entropy balance for the core:

\begin{equation}
\label{Eq-entropy}
\Phi + E_K = E_S + E_L + E_G.
\end{equation}

The right-hand side of Eq. (\ref{Eq-entropy}) contains the source
terms associated with secular cooling ($E_S$), release of latent heat
($E_L$), and release of gravitational potential energy ($E_G$).  
The sink term $E_K$ is the entropy associated with maintaining thermal conduction along an adiabatic temperature gradient.  
A convection-driven dynamo requires $\Phi>0$.  
Expressions for the terms of Eq. (\ref{Eq-entropy}) are listed in Table \ref{tab-terms} and are based on \cite{Nimmo2009} and \cite{Gaidos_etal2010} and references therein.

For a vigorously-convecting core one can assume a hydrostatic, isentropic, and well-mixed state.
The associated adiabatic density and temperature profiles can be approximated as
(\cite{Labrosse_etal2001,Labrosse2003,Nimmo2007}):
\begin{equation}
\label{Eq-density}
\rho(r) = \rho_{CMB} \exp { \left[\frac{r_{CMB}^2-r^2}{L^2} \right]}
\end{equation}
and 
\begin{equation}
\label{Eq-adiabat}
T_{ad}(r) = T_{CMB} \exp {\left( \frac{r_{CMB}^2(t) - r^2}{D^2}\right)},
\end{equation}
where $\rho_{CMB}$ and $T_{CMB}$ are the CMB density and temperature, respectively, and $L$ and $D$ are the density and thermal
length scales, respectively (see Table \ref{tab-notation}).  
The assumed thermal equilibration between inner and outer core is reasonable if the time scale of inner core convection is shorter than that of inner core growth \citep{Labrosse_etal2001}, a likely condition during the early stages of inner core growth \citep{Buffett2009}.

We calculated core solidification by expressing the solidus ($T_{sol}$) as a Lindemann law of melting (e.g., \cite{Labrosse_etal2001})
\begin{equation}
\label{Eq-solidus}
\frac{\partial \log T_{sol}}{\partial \log \rho} = -2\left( \gamma-\frac{1}{3} \right),
\end{equation}
where $\gamma$ is the Gr\"{u}neisen parameter.  We ignored the possible
effect of unknown concentrations of light elements such as sulfur on
the phase diagram of the core.  In all our calculations, the core
solidifies from the planetary center outward.  Solidification at the
top of the core (iron snow) is more likely for planets of two Earth
masses or more \citep{Gaidos_etal2010}.

We used the following energy balance equation for the core:
\begin{equation}
\label{Eq-energy}
Q_{CMB} =  Q_S + Q_L + Q_G.
\end{equation}
The source terms are the secular core cooling $Q_S$, the latent
heat release $Q_L$, and the gravitational energy release $Q_G$ (specified in
Table \ref{tab-terms}).  The CMB heat flow $Q_{CMB}$ is the only
relevant sink term and implicitly includes conductive heat transport
in the core along an adiabat.  We neglected core heating from the
decay of radioactive isotopes such as $^{40}$K, as well as viscous
dissipation \citep{BraginskyRoberts1995}.  Because $Q_S$, $Q_L$, and
$Q_G$ all depend linearly on the cooling rate $dT_{CMB}/dt$ (Table
\ref{tab-terms}), this cooling rate is readily solved if the heat flow
$Q_{CMB}$ is known.  
We use $dT_{CMB}/dt$ to solve the entropy terms (Eq. \ref{Eq-entropy}, Table \ref{tab-terms}) and to update the core temperature
profile (Eq. \ref{Eq-adiabat}) and inner core radius at model time steps of 1 Myr.

We characterized magnetodynamo evolution by 
(i) the onset time of inner solid core formation, $t_{ons}$, when $Q_L$ and $Q_G$ become nonzero and the dynamo intensifies, 
(ii) the dynamo cessation time, $t_{mag}$, defined as the time at which the dipole field intensity reduces to zero due to core freeze-out, and 
(iii) the maximum dipole field intensity from the start of the model until core solidification is complete, $B_{CMB}^{max}$.
We neglected the possibility that part of the liquid core would not contribute to the magnetodynamo due to stable thermal and/or chemical stratification.
However, we identified epochs of partial stabilization by calculating the onset times of thermal stratification, $t_{strat}$, i.e. when $Q_{CMB}<Q_K$ with $Q_K$ the thermal conduction along the adiabatic temperature gradient at the top of the core (Table \ref{tab-terms}).
For Earth, a stratified layer is probably $\lesssim$100 km thick \citep{Gubbins2007} and its influence on the magnetodynamo is likely small.

%-------------------------
\subsection{Dynamical model of the mantle}
\label{Sect-mantlemodel}

\begin{figure} 
%\noindent
\includegraphics[width=14cm]{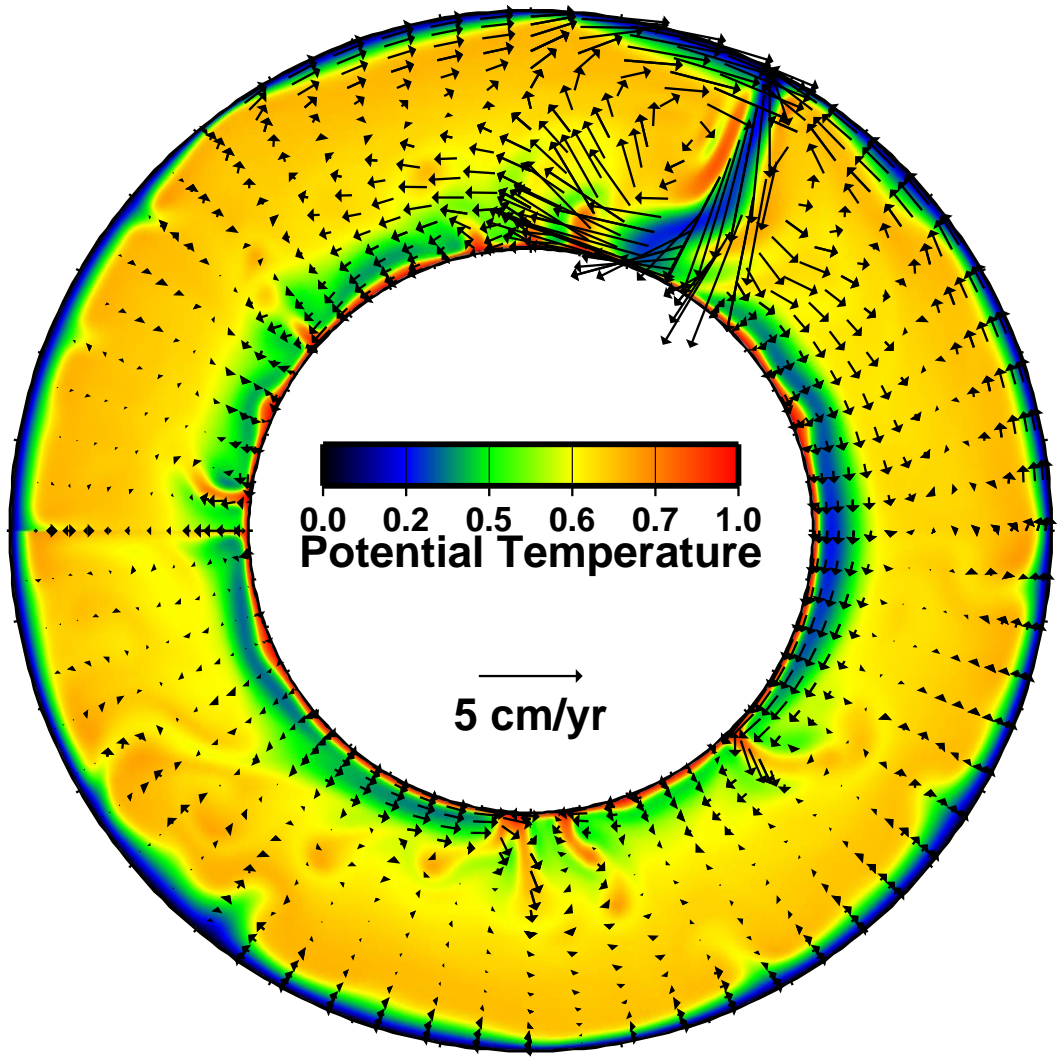}\\
\caption{
Convective flow field (arrows) and potential temperature (colors) at a model time of  $t=4.4$ Gyr in a 2-D, annular model of mantle convection for a nominal Earth-like scenario (C in Table \ref{tab-models}).
Dimensionless values of the mantle potential temperatures correspond to a range of 273--2673 K.
\label{fig-mantle} }
\end{figure}

We calculated $\overline{Q_{CMB}}$ values from numerical models of mantle convection with constant internal heat production ($4\times10^{-12}$ W kg$^{-1}$).
The methodology of our mantle convection models has been described in detail in \citet{vanSummeren_etal2011} and here we discuss only the aspects relevant for this paper.  
Mantle flow was
described by a Boussinesq formulation of an incompressible fluid at
infinite Prandtl number.  The associated conservation equations for
mass, momentum, and energy were solved using the finite element package
CitComS-3.1.1 \citep{Zhong_etal2000}.  The mantle domain is modeled as
a 360$^o$ annulus (Figure \ref{fig-mantle}), with 513 x 65 mesh nodes
in the lateral and vertical direction respectively, and mesh
refinement toward the top and bottom boundaries where the largest
viscosity contrasts occur.

To permit plate-like behavior of the surface boundary layer, we used a
composite viscous/pseudo-plastic rheology model \citep{Tackley2000}.
In low-stress regions, we assigned an Arrhenius-type rheology:
\begin{equation}
\label{Eq-viscosity}
\eta_v(T') = \eta_0 \exp \left[23.03 \left(\frac{1}{1+T'_p} - \frac{1}{2} \right) \right],
\end{equation}
with reference viscosity $\eta_0=5\times10^{20}$ Pa s.
The dimensionless potential temperature $T'_p$ scales to the dimensional potential temperature $T_p$ as $T'_p=(T_p-273)/2400$.
The viscosity $\eta_v$ decreases by 5 orders of magnitude when $T_p$ increases from 273 to 2673 K. 
In regions where the convective stress exceeds an assigned yield stress ($\sigma_y$), 
we implemented pseudo-plastic yielding by recalculating of the viscosity
$\eta_y=\sigma_y/2\dot{\epsilon_{II}}$, where $\dot{\epsilon_{II}}$ is the
second invariant of the strain rate tensor.  
This yielding allows for lithospheric break-up and concentrates strain in narrow regions.  
We computed the composite mantle viscosity as 
$\eta_m = \min\left(\eta_v(T'),\eta_y\right)$.  
We assigned $\sigma_y=150$ MPa in our convection experiments, 
except for the stagnant lid case where pseudo-plastic yielding is precluded.
This parameterization underestimates the temperature dependence of viscosity compared to experimental flow laws for olivine (e.g. \citep{HirthKohlstedt2003}), but it avoids sharp viscosity gradients that are computationally expensive.  
This viscosity parameterization is commonly used in modeling the dynamics of planetary mantles and successfully reproduces the phenomenology of plate-like tectonics by localizing strain in regions that resemble subduction zones and spreading ridges \citep{Tackley2000,VanHeckTackley2008,FoleyBecker2009,vanSummeren_etal2011}.
Although the viscous strength of plates is underestimated, plate tectonics is successfully reproduced because surface deformation occurs by yielding rather than viscous deformation.

For each scenario, we computed 8 cases with CMB potential temperatures in the range 
$T_{CMB_p}$=1873--3473 K ($T'_{CMB_p}$=0.67--1.33).  
We calculated physical (or ``real'')  CMB temperatures ($T_{CMB}$) from potential temperatures ($T_{CMB_p}$) by correcting for the influence of adiabatic compressibility: i.e., using 
$T_{CMB} = T_{CMB_p} \exp(Di)$, 
with 
dissipation number $Di=\alpha_m g h /c_{P_m}$ = 0.46,
mantle thermal expansivity  $\alpha_m = 2\times10^{-5}$ K$^{-1}$, 
mantle gravitational acceleration $g$ = 10 m s$^{-2}$,
mantle thickness $h=2.891\times10^6$ m,
and mantle heat expansivity $c_{P_m}=1250$ J kg$^{-1}$ K$^{-1}$.
Because each model has a constant CMB temperature and heat production, $Q_{CMB}$ eventually reaches a statistically steady state value, $\overline{Q_{CMB}}$.  
Typically, we ran each case for a model time of $\sim$9 Gyr (i.e. 32,000 model time steps) and we removed the first $\sim$5 Gyr (12,000 time steps) to exclude initial transients.
For the results presented in Section \ref{Sect-res-life} we increased the maximum core temperature to $T_{CMB}^{init}$ = 5000 K by extrapolating the heat flow data, using linear regression to the filled symbols shown in Figure \ref{fig-QT}b. 
We applied the same technique to estimate $Q_{CMB}$ for $T_{CMB}>$4500 K for scenario VI (i.e. the case represented by the right-hand brown circle in Figure \ref{fig-QT}a relies on extrapolation). 

\begin{figure} 
\includegraphics[width=9cm]{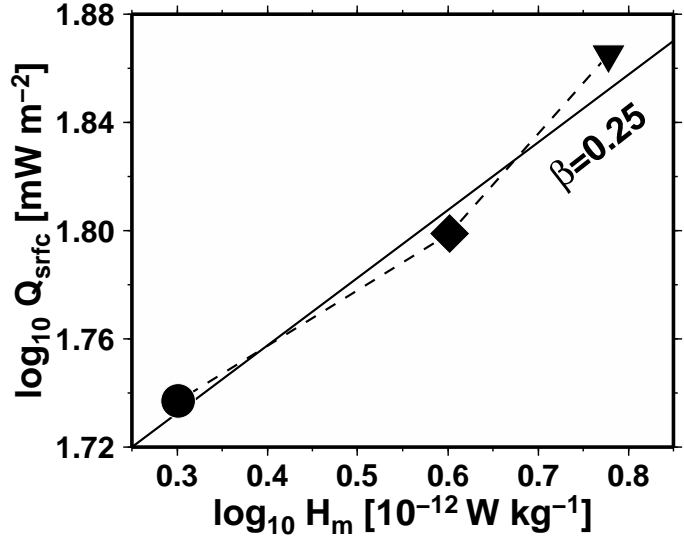}\\
\caption{
Statistically steady state surface heat flow for models with (time-independent) mantle internal heating $H_m$=2, 4, and 6 $\times10^{-12}$ W kg$^{-1}$ (circle, diamond, and triangle, respectively).
The model with $H_m=4\times10^{-12}$ W kg$^{-1}$ (diamond) is reference scenario C (Figure \ref{fig-mantle}, Table \ref{tab-models}).
The solid line shows the theoretical trend from boundary layer theory for internally heated systems ($Q_{srfc}=H_m^{0.25}$).
\label{fig-qh} }
\end{figure}

To account for a time-dependent rate of mantle internal heat production, we rescaled the CMB heat flow using the quasi steady state condition $Q_{CMB}(t) = Q_{srfc}(t) - H_m(t)$.
We scaled the surface heat flow by $Q_{srfc}(t) \propto Ra^{\beta}_H(t)$, 
with $Ra_H(t)=\alpha_m \rho_m^2 g H_m(t) h^3 / k_m \kappa_m\eta_m$ the Rayleigh number for an internally heated fluid,
$k_m$=5.65 W$^{-1}$ K$^{-1}$ the mantle thermal conductivity, 
and $\rho_m = 4500$ kg m$^{-3}$ the mantle reference density. 
We thus have $Q_{srfc}(t) \propto H^{\beta}_m(t)$.
We also assumed a quasi steady state heat balance for the unscaled terms (indicated with subscripts ``0''): $Q_{srfc_0}=Q_{CMB_0}+H_{m_0}$. 
We then rescaled the CMB heat flow with
\begin{equation} 
Q_{CMB}(t)=\left( Q_{CMB_0}+H_{m_0} \right) \left[ \frac{H_m(t)}{H_{m_0}} \right] ^{\beta} - H_m(t)
\end{equation} 
We used $\beta=0.25$, consistent with boundary layer theory for an internally heated mantle.
Deviations from the theory may arise, for example due to contributions from bottom heating or the dependence of $Ra_H$ on mantle heating rate through $\eta_m$.
We validated that $\beta$=0.25 is appropriate for our models by calculating the surface heat flow from two verification experiments with mantle internal heating rates $2\times10^{-12}$ and $6\times10^{-12}$ W kg$^{-1}$ (Figure \ref{fig-qh}), which yielded $\beta=0.27$. 
We evolved the mantle heat production rate with respect to time as:
\begin{equation}
H_m(t) = \sum_{i} C^i H^i \exp\left[\frac{-t \ln 2}{\tau_{1/2}^i}\right], 
\label{Eq-heatprod}
\end{equation}
with isotope concentration $C$, heat production rate $H$, and half-life time $\tau_{1/2}$, for radiogenic isotopes $^{238}$U, $^{235}$U, $^{232}$Th, and $^{40}$K (values from \citet{TurcotteSchubert2002}, Table 4-2).
Estimates for present-day mantle heating (e.g. $7.4\times10^{-12}$ W kg$^{-1}$, \citep{TurcotteSchubert2002}) depend on assumptions of the Urey number.
To account for this uncertainty, we considered a range of $H_{m_0}$ values (1$\times10^{-12}$ -- 8$\times10^{-12}$ W kg$^{-1}$) and adjusted the heat production rates for radiogenic isotopes $H^i$ (Eq. \ref{Eq-heatprod}), accordingly.

For comparison with boundary layer theory, we recast our results in terms of dimensionless parameters and express $Q_{CMB}$ as
\begin{equation}
\label{Eq-Nu}
Q_{CMB} = 4 \pi  r_{CMB} k_m (T_{CMB_p}-<T_{m_p}>) Nu_{CMB},
\end{equation}
with $Nu_{CMB}$ the Nusselt number for the bottom boundary layer.
Boundary layer theory predicts $Nu_{CMB} \propto Ra_{CMB}^{1/3}$ for a
bottom-heated mantle \citep{Schubert_etal2001}.  We define the
Rayleigh number for the bottom boundary layer as
\begin{equation}
\label{Eq-Ra}
Ra_{CMB} = \frac{\rho_m g \alpha_m (T_{CMB_p} - <T_{m_p}>)h^3} {\kappa_m <\eta_m>}.
\end{equation}

%-------------------------
\subsection{Parameter values and model scenarios}
\label{Sect-cases}

We calculated CMB heat flows for 5 scenarios of mantle convection and temperature (Table \ref{tab-models}). 
We simulated a nominal Earth (``cold'' or C model) with a uniform surface temperature $T_{srfc}$ = 273 K.
Using a ``hot'' (H) model with a uniform $T_{srfc}$ = 759 K we simulated a close-in planet ($a$=0.13 AU) with efficient surface heat redistribution.  
An ``asymmetric'' (A) model was used for close-in planets ($a$=0.13 AU) that lack heat redistribution, and we assigned a surface temperature that is elevated at the substellar point ($T_{srfc}$=1073 K), decreases gradually to the terminus, and is constant on the night side ($T_{srfc}$ = 273 K).  
To investigate planets with an immobile surface and convecting interior, such as Venus, we ran a ``stagnant lid'' (SL) model by removing pseudo-plastic behavior from the viscosity parameterization.
For the ``viscosity increase'' (VI) model we assigned an exponential increase of the viscosity by a factor of 10 across the mantle depth range, in addition to temperature-dependent viscosity (Eq. \ref{Eq-viscosity}).

We complemented our investigation of different convection regimes (Sections \ref{Sect-QT}--\ref{Sect-MD}) with a study of the sensitivity of results to core properties (Section \ref{Sect-res-life}).
We calculated the sensitivity of the dynamo cessation time ($t_{mag}$) and maximum CMB field intensity ($B_{CMB}^{max}$) to two planetary properties: core radius ($r_{CMB}$) and initial CMB temperatures ($T_{CMB}^{init}$), 
all for the nominal C scenario of mantle tectonics
We also calculated the sensitivity of our results to core solidus ($T_{sol}$) and heat conductivity ($k_{c}$), parameters for which the values at pressures corresponding to Earth's center are controversial \citep{Morard_etal2011}.  
Our solidus range equates to a range of the
Gr\"{u}neisen parameter $\gamma \in$ [1.1--1.4], reflecting current uncertainties in high-pressure mineral physics.  
Where we varied $r_{CMB}$, we recalculated the interior structure using the interior model described in \citet{Gaidos_etal2010}.
This uses the third-order Birch-Murnagham (BM) equations of state, and includes the Thomas-Fermi-Dirac contribution to the pressure using the formulation of \citet{Zapolsky69}.  
The density in the liquid part of the core is adjusted by a fixed fraction $\delta \rho/\rho$ to account for the presence of light elements \citep{Lee2003}.  
We adjusted the solidus at the planetary center according to Eq. (\ref{Eq-solidus}).  
For a valid comparison we adjusted the mantle thickness to maintain a constant Rayleigh number, i.e. the product of mantle density, gravitational acceleration, and the cube of the mantle thickness, evaluated at the CMB.

%-------------------------
\section{Results}
\subsection{Mantle dynamics and CMB heat flow}
\label{Sect-QT}

\begin{figure} 
%\noindent
\includegraphics[width=14cm]{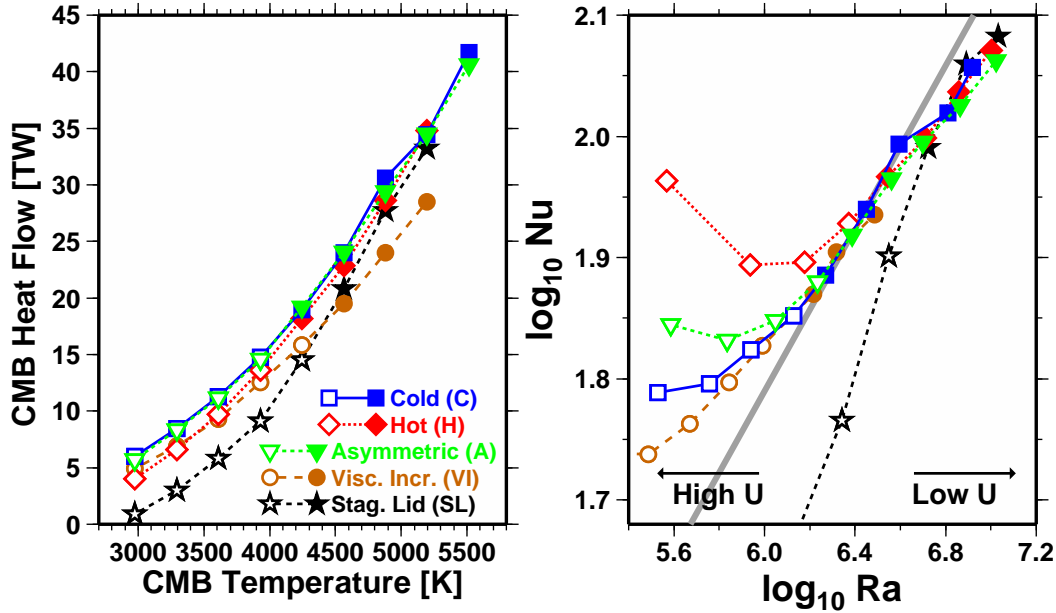}\\
\caption{
(a) Statistically steady state CMB heat flow, $\overline{Q_{CMB}}$,  as a function of the CMB temperature, $T_{CMB}$.  
Results are for mantle convection scenarios C (cold surface), H (hot surface), A (surface with asymmetric temperature), SL (stagnant lid), and VI (depth-increasing viscosity) (see Table \ref{tab-models}). 
Filled symbols indicate cases where bottom heating dominates over internal heat production in the mantle (U$<$0.5) (b) Heat flow curves recast in terms of dimensionless parameters $Nu_{CMB}$ (Eq. \ref{Eq-Nu}) and $Ra_{CMB}$ (Eq. \ref{Eq-Ra}).  
The grey line depicts the power-law relationship $Nu_{CMB}\propto Ra^{1/3}$ predicted from boundary layer theory.  
  \label{fig-QT} }
\end{figure}

Figure \ref{fig-QT}a shows our computed CMB heat flow $\overline{Q_{CMB}}$ as a function of $T_{CMB}$ for the 5 mantle convection scenarios described in Section \ref{Sect-mantlemodel} and listed in Table \ref{tab-models}.
Models C, H, and A build on the work of \citet{vanSummeren_etal2011} and their dynamics will be discussed only briefly.  
The nominal Earth-like scenario C with a cold surface ($T_{srfc}$=273 K) is characterized by plate-like behavior with rigid lithospheric plates that are broken by narrow regions of high strain that resemble slab-like downwellings and ridge-like upwellings (Figure \ref{fig-mantle}).  
The $\overline{Q_{CMB}}$ values increase with increasing $T_{CMB}$ (Figure \ref{fig-QT}a, blue curve), which reflects elevated heat transport in a more vigorously convecting mantle.

In scenario H of a close-in planet with a uniformly elevated surface temperature ($T_{srfc}$=759 K), the surface boundary is too weak to maintain coherent plates.  
Instead, the surface material is highly mobile and deforms diffusely,
a possibility for close-in planets with surface temperatures close to the silicate solidus.
We limited the temperature dependence of viscosity for numerical reasons and therefore likely underestimated the viscosity of the surface boundary layer.
Planets with $T_{srfc}$=759 K may instead have 
immobile or episodically overturning surfaces \citep{ArmannTackley2012},
 and we investigated this stagnant lid behavior with our SL models, described below. 
For scenario H, the $\overline{Q_{CMB}}$ values are only slightly smaller than for scenario C (Figure \ref{fig-QT}a, cf. red and blue curves).
Indeed, the elevated surface temperature causes a mantle temperature increase that reduces the CMB heat flow, but this is mostly offset by a lower viscosity in the hot boundary layer that enhances CMB heat flow.

In scenario A, the hemispheric contrast in surface temperatures causes a tectonic dichotomy with diffuse deformation on the hot day side (similar to scenario H) and plate-like tectonics on the cold night side (similar to scenario C) \citep{vanSummeren_etal2011}.
This mixed style of surface deformation results in $\overline{Q_{CMB}}$ values that are intermediate those of scenarios C and H (Figure \ref{fig-QT}a).
An asymmetric (harmonic degree 1) convection pattern develops in the mantle interior.
This pattern involves convective upwellings that are concentrated near the hot sub-stellar point, near-surface flow from the hot day side to the cold night side where downwellings concentrate, and a deep mantle return flow toward the day side.
A small hemispheric heterogeneity in CMB heat flow of $\sim$10\% in scenario A is unlikely to destabilize a dynamo \citep{OlsonChristensen2002}. 

In scenario VI (in which viscosity increases with pressure) CMB heat flows are $\sim$15-25\% lower than in scenario C (Figure \ref{fig-QT}a).
This reflects the depth-increasing viscosity which suppresses deep mantle convective overturn and hinders descending cold slabs from covering and cooling the core.
As a result, a thicker, less conducting CMB boundary layer develops, relative to reference scenario C.

The stagnant lid convection scenario SL has an immobile surface through which heat is conducted, but vigorous convection still occurs in the underlying mantle.  
The $\overline{Q_{CMB}}$ values are lower than for models C, H, and A for the $T_{CMB}$ range we investigated (Figure \ref{fig-QT}b).
This is because slow heat transport through the immobile surface reduces the temperature difference between the mantle interior and the core.

We expressed our results in dimensionless quantities, i.e. the Nusselt and Rayleigh numbers of the bottom boundary layer as defined by Eq. (\ref{Eq-Nu}) and Eq. (\ref{Eq-Ra}).  
When bottom heating dominates (i.e. for Urey number $U< $0.5), our results approach the theoretical 1/3 power-law relationship for a bottom-heated mantle \citep{Schubert_etal2001} (Figure \ref{fig-QT}b).
Calculated $Nu_{CMB}$ values deviate from the theoretical value when internal heating dominates ($U> $0.5), which occurs for $T_{CMB}\lesssim$4000--4500 K in our models (Figure \ref{fig-QT}b).
For $U>$0.5, $Nu_{CMB}$ is larger for scenarios H and A than for scenario C (Figure \ref{fig-QT}b) because (locally) elevated surface temperatures thermally buffer the mantle interior and reduce the temperature contrast with the core (Eq. \ref{Eq-Nu}).
The CMB heat flow also decreases, but this is mostly compensated by more efficient heat transport resulting from decreased mantle viscosity.
Compared to scenario C, scenarios VI and SL have a lower $Nu_{CMB}$ which reflects a lower CMB heat flow due to the weaker influence of cold slabs in the deep mantle.

%-------------------------
\subsection{Nominal thermal evolution of the core}

\begin{figure} 
\includegraphics[width=11cm]{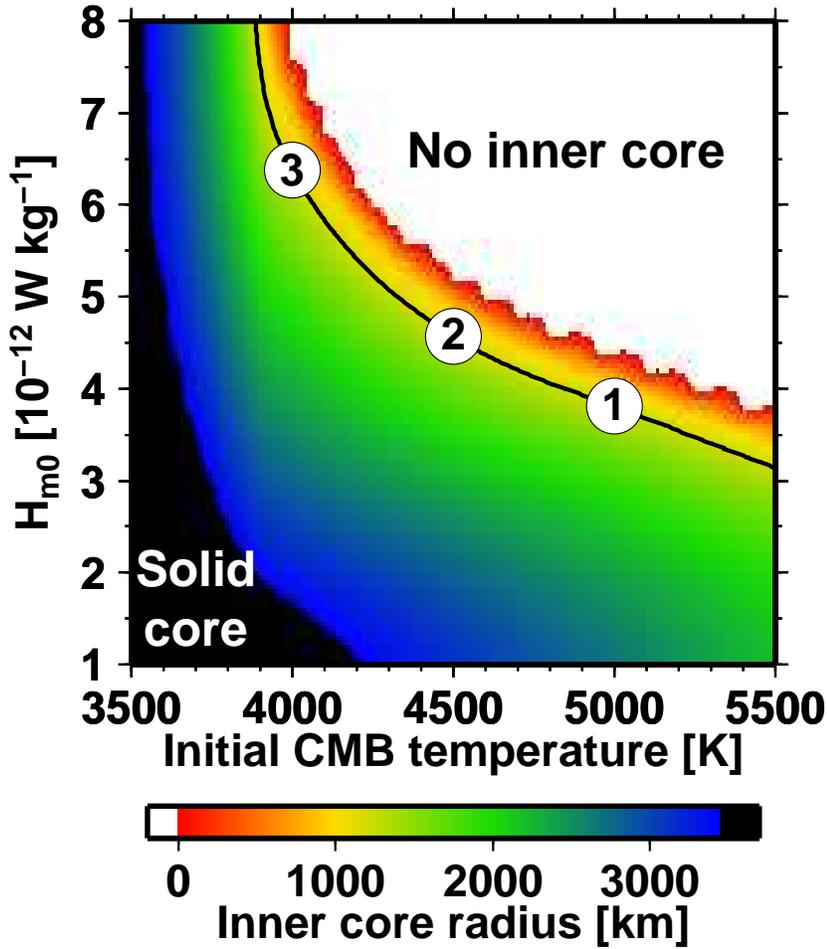}\\
\caption{
Calculated inner core radius after a model time of $t=$4.5 Gyr as a function of initial CMB temperature ($T_{CMB}^{init}$) and present-day radiogenic mantle heat production rate ($H_{m_0}$).
The black contour marks Earth's present-day inner core radius of 1220 km. 
Labels indicate 3 scenarios that reproduce a 1220 km inner core at $t$=4.5 Gyr by using respective values of $H_{m_0}$ and $T_{CMB}^{init}$:
(1) 3.8$\times10^{-12}$ W kg$^{-1}$, 5000 K; 
(2) 4.6$\times10^{-12}$ W kg$^{-1}$, 4500 K;
(3) 6.4$\times10^{-12}$ W kg$^{-1}$, 4000 K. 
Corresponding dynamo evolutions for these cases are shown in Figure \ref{fig-corevol}d.
\label{fig-icradius} }
\end{figure}

\begin{figure} 
\includegraphics[width=12cm]{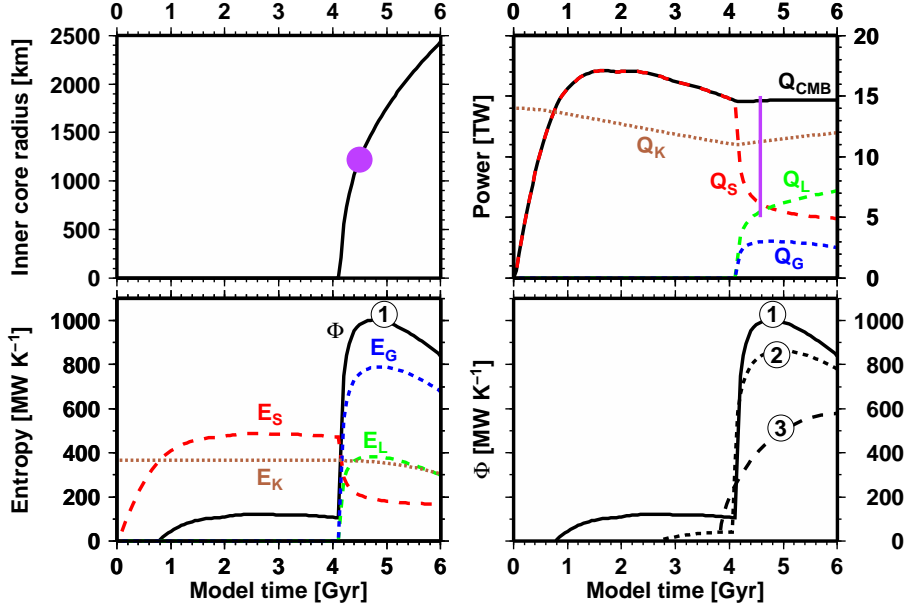}\\
\caption{
Thermal evolution of the core for nominal Earth-like scenario C. 
(a) Inner core radius.
The purple circle shows Earth's present-day value.
(b) Energy and (c) entropy terms (Eqs. (\ref{Eq-entropy}) and (\ref{Eq-energy}) and Table \ref{tab-terms}).
Label subscripts K, L, S, and G are for conduction, latent heat, secular cooling and graviational energy release, respectively.
$Q_{CMB}$ is the CMB heat flow and $\Phi$ is the entropy production rate associated with ohmic dissipation.
The initial CMB temperature used in this model is $T_{CMB}^{init}$=5000 K and the present-day mantle heat production rate is $H_{m_0}=3.8\times10^{-12}$ W kg$^{-1}$ (case 1 in Figure \ref{fig-icradius}).
The purple bar in (b) indicates  a range of recent CMB heat flow estimates for Earth \citep{Lay_etal2008}. 
(d) Evolution of $\Phi$ for 3 models that reproduce Earth's present-day inner core radius for $H_{m_0}$ and $T_{CMB}^{init}$ values as indicated by the corresponding points labeled (1-3) in Figure \ref{fig-icradius}.
\label{fig-corevol} }
\end{figure}

We calibrated the nominal Earth-like scenario by reproducing Earth's present-day inner core radius of 1220 km
and heat flow of 5-15 TW \citep{Lay_etal2008})
at a model time of 4.5 Gyr.
In calculating the thermal evolution of the core, we accounted for time-dependent mantle heating by rescaling the parameterized CMB heat flow values (Figure \ref{fig-QT}, blue curve) and investigated the range of $T_{CMB}^{init}$ and $H_{m_0}$ values shown in Figure \ref{fig-icradius}.
Different combinations of parameter values can reproduce Earth's inner core radius.
For example, the calculated present-day inner core radius increases for progressively lower $T_{CMB}$, lower $H_{m_0}$, or higher solidus $T_{sol}$ (i.e., larger Gr\"{u}neisen parameter $\gamma$) and the present-day $Q_{CMB}$ increases for higher $T_{CMB}$, lower $H_{m_0}$, or higher $T_{sol}$. 
We chose $T_{CMB}^{init}$=5000 K, $H_{m_0}$=3.8$\times10^{-12}$W kg$^{-1}$ (label 1 in Figure \ref{fig-icradius}), and $\gamma$=1.3 ($T_{sol}^{cen}$=5391 K) and
the corresponding core evolution (Figures \ref{fig-corevol}a-c) serves as a reference for other scenarios. 
Dynamo operation starts at $t\sim$0.8 Gyr ($\Phi>0$, Figure \ref{fig-corevol}c), when the CMB heat flow has become sufficiently large ($\sim$14 TW) after an initial period of strong mantle heating that suppresses the CMB heat flow (Figure \ref{fig-corevol}b).
After inner core formation ($\sim$4.1 Gyr), latent heat and gravitational energy release contribute to the entropy production in the core,
which causes an abrupt increase in $\Phi$ (Figure \ref{fig-corevol}b) and a stronger dynamo. 
As the core continues to solidify, $\Phi$ gradually decreases due to the progressive thinning of the fluid iron layer.
Finally, dynamo operation ceases at $\sim$8.2 Gyr when core solidification is complete.
The onset of inner core formation coincides with a change in the $Q_{CMB}$ slope from negative to nearly flat (Figure \ref{fig-corevol}b, $\sim$4.1 Gyr).
This is due to 
(i) the generation of latent heat and gravitational potential energy, 
which partly replenishes heat carried from the core and maintains high core temperatures, and
(ii) cooling of the mantle due to decreasing internal heat production, which acts to enhance $Q_{CMB}$.

We calculated the sensitivity of dynamo evolution to $H_{m_0}$ and $T_{CMB}^{init}$ using 3 combinations that reproduce Earth's present-day inner core radius (values as specified in Figure \ref{fig-icradius}).
A progressive decrease of $T_{CMB}^{init}$ by 500 K lowers the  CMB heat flow, which is reflected in a corresponding decrease in $\Phi$ and a delay in the onset time of dynamo operation (Figures \ref{fig-icradius} and \ref{fig-corevol}d, labels 1-3).
Among the models that reproduce Earth's present-day inner core radius, only those with $T_{CMB}^{init}\gtrsim5000$ K correspond with an early non-magnetic epoch $<$1 Gyr (label 1 in Figures \ref{fig-icradius} and \ref{fig-corevol}d), in agreement with evidence for Earth's long-lived ($>$3.5 Ga) magnetic field \citep{Biggin_etal2008}.
We therefore prefer case 1 and use the corresponding values in subsequent models, unless otherwise mentioned.

%-------------------------
\subsection{Dynamo evolution for different mantle convection regimes}
\label{Sect-MD}

\begin{figure} 
%\noindent
\includegraphics[width=9cm]{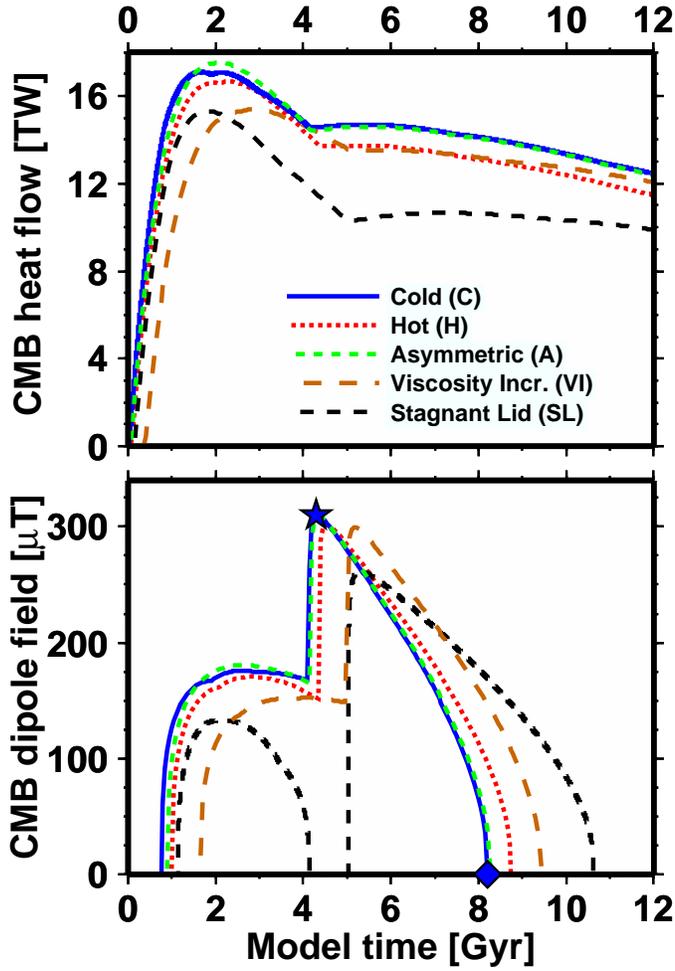}\\
\caption{
Dynamo evolution for 5 scenarios of mantle convection (Section \ref{Sect-cases}, Table \ref{tab-models}).
(a) Heat flow at the CMB, $Q_{CMB}$. 
Time-dependent mantle heat production is taken into account by rescaling of the steady state $\overline{Q_{CMB}}$ vs. $T_{CMB}$ parameterizations shown in Figure \ref{fig-QT}a. 
(b) Magnetic dipole field intensity calculated at the CMB, $B_{CMB}$.
The star and diamond indicate maximum CMB field intensity, $B_{CMB}^{max}$, and time of dynamo cessation, $t_{mag}$, for the nominal Earth-like scenario C.
The respective symbols correspond with those in Figure \ref{fig-lifetime}. 
\label{fig-MD} }
\end{figure}

Figure \ref{fig-MD} shows the evolution of the dynamo in the 5 mantle convection scenarios (Section \ref{Sect-cases}, Table \ref{tab-models}).
Throughout most of the evolution, the CMB heat flows for scenarios with plate tectonics or diffuse surface deformation (C, H, A, VI) are larger by $\sim$2--4 TW than for the stagnant lid scenario SL (Figure \ref{fig-MD}a),  
a consequence of the $\overline{Q_{CMB}}$ vs. $T_{CMB}$ systematics (Figure \ref{fig-QT}).
The impact of these contrasting heat flows on dynamo evolution is shown in Figure \ref{fig-MD}b.
 For scenarios with active surface recycling (C, H, A, and SL), dynamo operation starts at $\sim$0.8--1.6 Gyr and continues until core freeze-out ($\sim$8.2--9.5 Gyr). 
For the stagnant lid scenario SL, an early dynamo ceases at $\sim$4.1 Gyr because of the low CMB heat flow and corresponding low entropy production in the core.
After a magnetic quiet period ($\sim$4.1--5.0 Gyr), the dynamo can restart due to inner core formation and continues until $\sim$10.6 Gyr, i.e. longer than for all other scenarios that we investigated.
The longevity of this second-stage dynamo is due to late freeze-out in a slowly cooling single-plate planet with relatively low CMB heat flow.
Unlike the other scenarios, thermal stratification can occur during a substantial part of the evolution of this core ($t_{strat}$=4.4 Gyr, Table \ref{tab-models}). 
The CMB dipole field strengths $B_{CMB}$ are of similar magnitude in the 5 different scenarios due to the weak dependence of $B_{CMB}$ on $\Phi$ (Eq. \ref{Eq-scalinglaw}) and this is reflected in the limited variation in the maximum field intensities in these scenarios ($B_{CMB}^{max}$=260--310 $\mu$T, Table \ref{tab-models}).

%-------------------------
\subsection{Sensitivity of dynamo evolution to core properties}
\label{Sect-res-life}

\begin{figure} 
%\noindent
\includegraphics[width=14cm]{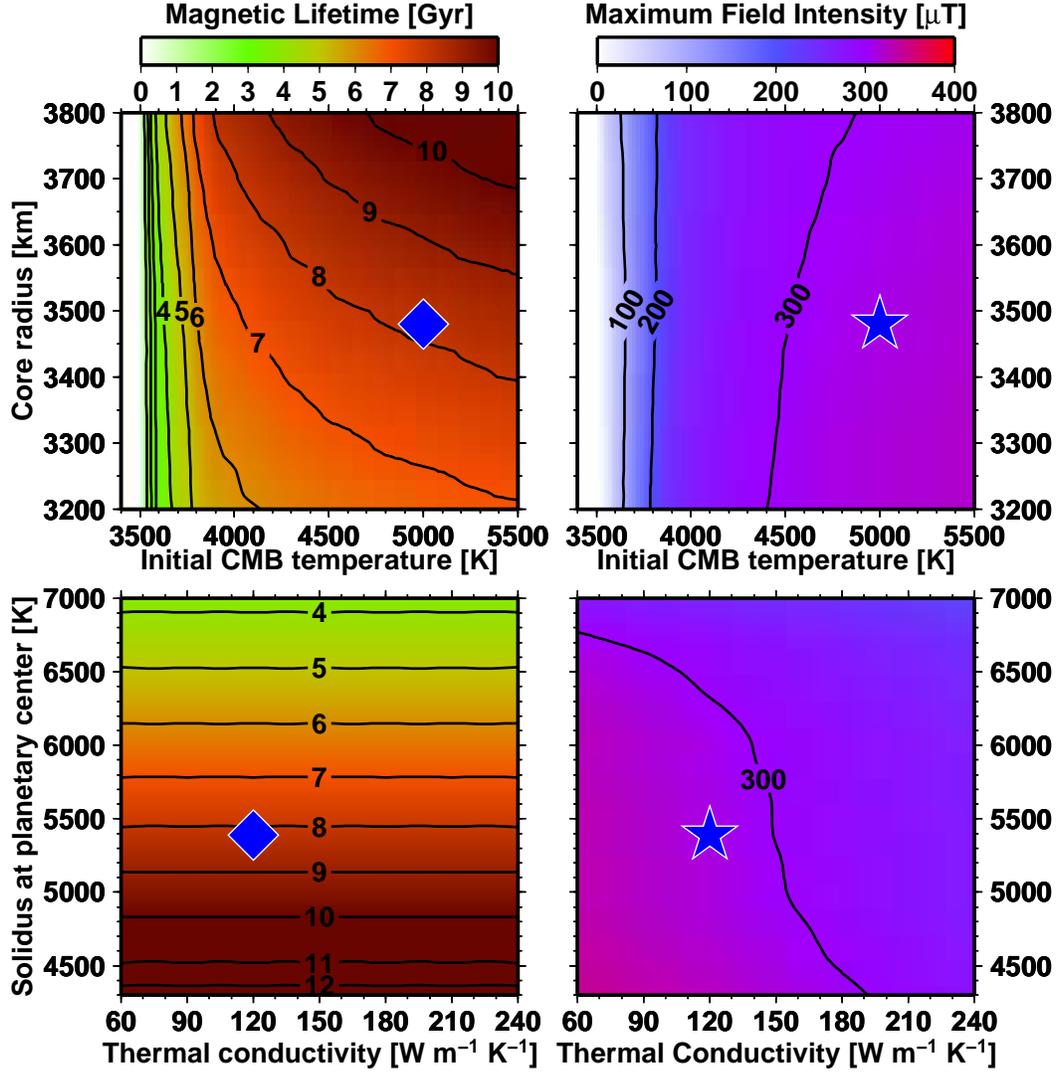}\\
\caption{
Sensitivity of (a and c) dynamo cessation time $t_{mag}$, and (b and d) maximum dipole field intensity at the CMB, $B_{CMB}^{max}$, as a function of (a and b) initial CMB temperature $T_{CMB}^{init}$ and core radius $r_{CMB}$, and (c and d) core heat capacity $k_c$ and solidus temperature at the planet's center $T_{sol}$.  
Stars and diamonds respectively indicate $B_{CMB}^{max}$ and $t_{mag}$ for reference scenario C (Table \ref{tab-models}) and correspond with the symbols in Figure \ref{fig-MD}b .
\label{fig-lifetime} }
\end{figure}

We investigated the sensitivity of dynamo evolution to core radius $r_{CMB}$, initial temperature $T_{CMB}^{init}$, and material properties (solidus $T_{sol}$ and thermal conductivity $k_{c}$) in the nominal Earth-like scenario.
Figures \ref{fig-lifetime}a and b show that hotter and bigger cores prolong dynamo operation.
Planets with $T_{CMB}\lesssim$3500 K are non-magnetic because they (unrealistically) start with and retain a solid core.
A progressive increase in $T_{CMB}^{init}$ causes a subsequently smaller increase of dynamo cessation times (Figure \ref{fig-lifetime}a) because hotter cores lose their heat more efficiently (Figure \ref{fig-QT}a). 
 
Recent first principle computations of liquid iron mixtures \citep{DeKoker_etal2012,Pozzo_etal2012} suggest a core thermal core conductivity 2--3 times higher than previous estimates that relied on extrapolations, e.g. \citet{StaceyAnderson2001}. 
Our results show that dynamo cessation times $t_{mag}$ are insensitive to $k_{c}$ in the range 60-240 W m$^{-1}$ K$^{-1}$ (Figures \ref{fig-lifetime}c and d).
An increase in $k_c$ increases the entropy term related to heat conduction $E_K$ and this weakens the dynamo (lowers $\Phi$, Eq. \ref{Eq-entropy}). 
For $k_c\gtrsim$160 W m$^{-1}$ K$^{-1}$, the dynamo collapses prior to inner core formation, although this does not affect the cessation times $t_{mag}$ (Section \ref{Sect-coremodel}).
Dynamo cessation after inner core formation requires even higher $k_c$ ($\gtrsim$500 W m$^{-1}$ K$ ^{-1}$, outside the range shown in Figure \ref{fig-lifetime}c) to offset the latent heat and gravitational energy release contributions.
Dynamo operation is insensitive to $k_c$ after inner core formation and for  $k_c\lesssim$500 W m$^{-1}$ K$^{-1}$ because: 
(i) the time of core freeze-out is controlled by $Q_{CMB}$, which depends on the convective vigor of the mantle but is independent of $k_c$,
(ii) the entropy sink term $E_K$ reduces to zero and this 
removes the dependence of $\Phi$ on $k_c$ (Table \ref{tab-terms}).
These $k_c$ limits are derived using a central solidus $T_{sol}$=5391 K.
Dynamo lifetimes decrease for progressively larger $T_{sol}$ (Figures \ref{fig-lifetime}c and d) because planetary cores solidify at higher temperatures, i.e. generally earlier in the planet's history.

Peak field intensities are only weakly sensitive to core properties ($r_{CMB}$, $T_{CMB}^{init}$, $T_{sol}$, $k_{c}$) because maximum intensities occur at inner core formation (Figures \ref{fig-corevol}c and \ref{fig-MD}c) which occurs at a specific temperature regardless of the onset time and because $B_{CMB}\sim \Phi^{1/3}$ (Eq. \ref{Eq-scalinglaw}).
Intensities decrease sharply, however, when dynamo lifetimes reduce to zero, 
i.e. for low initial core temperatures ($T_{CMB}^{init}\lesssim$3500 K) and high solidus ($T_{sol}^{cen}\gtrsim$8400, Figures \ref{fig-lifetime}b and d, solidus limit out of the range shown).

%%%%%%%%%%%%%%%%%%%%%%%%%%%%%%%%%%%%%%%%
\section{Discussion}

%-----------------------------
\subsection{Comparison with Earth and Venus}
\label{Sect-EarthVenus}

Our nominal Earth-like scenario successfully reproduces evidence for a geodynamo that has operated since at least 3.5 Ga \citep{Biggin_etal2008}, and Earth's present-day inner core radius of 1220 km.
The corresponding present-day CMB temperature of 3875 K in our models is compatible with experimental and computational estimates for iron alloys of 3700--4300 K (e.g., \citet{Boehler2000}) and estimates of 3600--4200 K based on seismologically inferred deep mantle structures combined with calculated elastic properties of lower mantle minerals \citep{Lay_etal2006,VanderHilst_etal2007,KawaiTsuchiya2009}. 
These constraints restrict the evolution of core energetics, which critically depends on the CMB heat flow.
In our preferred scenario, the CMB heat flow is $\sim$15 TW at $t$=4.5 Gyr, consistent with recent estimates of the present-day CMB heat flow (5--15 TW, \citet{Lay_etal2008}).
The CMB heat flow varies between $\sim$14 and 17 TW, except for the first $\sim$1 Gyr when strong mantle heat production impedes core cooling. 
Our calculations do not require a source of radiogenic heat in the core to match Earth's long-lived dynamo, 

A relatively young inner core ($\sim$500 Ma) and a maximum rate of entropy production associated with ohmic dissipation before and after inner core formation of respectively $\Phi \sim$100 MW K$^{-1}$ and $\sim$1000 MW K$^{-1}$ are consistent with other core evolution studies with comparable CMB heat flow values \citep{Labrosse2003,Nimmo2007}. 
Previous studies have computed older cores ($\sim$1 Ga), likely because of a lower $Q_{CMB}\sim$10 TW \citep{Labrosse2003,Breuer_etal2010,Gaidos_etal2010}.
We prefer relatively high values that prevent the formation of a stably stratified layer in the liquid core (requiring $Q_{CMB}>Q_K\sim$11--14 TW, Figure \ref{fig-corevol}b) which, for Earth, is absent or thin ($\lesssim$100 km, \citep{Gubbins2007}).

In this light, additional mechanisms that can modulate Earth's CMB heat flow are important. 
Enhanced CMB heat flow is possible from the decay of $^{40}$K in Earth's core, and destabilization of the mantle bottom boundary layer due to a viscosity reduction within deep mantle post-perovskite regions 
\citep{NakagawaTackley2011}.
It is possible that Earth's viscosity increase with depth is larger (factor of $\sim$1000, \citet{MitrovicaForte2004,SteinbergerCalderwood2005}) than employed in our models.
Our results show that an increase of the pressure dependence reduces convective overturn in the deep mantle, which lowers the CMB heat flow and weakens the dynamo (cf. models C and VI, Figure \ref{fig-MD}), consistent with parameterized convection models \citep{Tachinami2011}.
Dynamo operation may be viable for planets with strong pressure-dependence, however, because of sustained convective heat loss due to feedback between viscosity, temperature, and internal heating \citep{Tozer1965,Tackley2012}.
Possible mechanisms that can decrease the CMB heat flow include a reduced radiative conductivity due to high-pressure transitions from high-spin to low-spin in iron atoms \citep{Badro_etal2004,Goncharov_etal2006}, thermal insulation due to a compositionally-dense layer in the deep mantle \citep{NakagawaTackley2004,NakagawaTackley2010}, and reduced convective heat transport due to a decrease of the thermal expansivity with pressure \citep{ChopelasBoehler1992}.

It has been widely debated why a dynamo is absent on Venus, which has a similar size and interior structure as Earth but lacks plate tectonics and has a higher surface temperature.
Our stagnant lid model results raise the possibility that a dynamo is currently inactive in Venus because of a low CMB heat flow in the absence of a crystallizing inner core.
If true, a dynamo may restart after the onset of core freezing which enhances the entropy production.
This scenario is consistent with previous calculations of \citet{Stevenson_etal1983} and would not require 
additional mechanisms, such as reduced mantle convection due to dehydration stiffening \citep{NimmoMcKenzie1996,Nimmo2002a,Gaidos_etal2010}.
Indeed, the influence of an intrinsic viscosity increase may be limited by strong viscosity-temperature feedback \citep{Tozer1965}.

%--------------------------
\subsection{Uncertainties and model approximations}

Several uncertainties and model approximations temper strong statements about dynamo operation in rocky planets.
Our viscosity parameterization underestimates the temperature dependence of viscosity of mantle silicates compared to laboratory experiments.
Planetary mantles with a stronger temperature dependence likely experience a smaller temperature decrease with time because thermal convection tends to be more self-regulating \citep{Tozer1965}.
If this causes a smaller temperature difference with the core, a lower CMB heat flow is expected to weaken dynamo operation. 
On the other hand, 
the bottom thermal boundary layer would have a lower viscosity, which locally enhances the mobility of upwellings and this could increase the CMB heat flow and  produce a stronger, but shorter-lived dynamo.
For planets with immobile surfaces, \citet{LiKiefer2007} showed that a stronger temperature dependence results in a thicker upper boundary layer and this reduces the surface heat flow and increases mantle temperatures, which reduces the CMB heat flow.
A reduced CMB heat flow likely delays inner core formation and hampers dynamo operation until an inner core forms and this strengthens our explanation for a present-day non-magnetic Venus. 

In rescaling the CMB heat flow to account for time-dependent mantle heating we have neglected secular cooling of the mantle.
Secular mantle cooling would contribute to the energy balance in a similar way as the radiogenic mantle heat production. 
We may therefore overestimate the CMB heat flow and overestimate dynamo intensities although, for Earth, secular cooling is likely $\sim$3 times smaller than internal heating \citep{TurcotteSchubert2002}.  
However, a larger early CMB heat flow would develop if the early core was super-heated relative to the mantle during an out-of-equilibrium thermal state immediately following core-mantle differentiation \citep{Stevenson1990}.
Also, magmatism, while not included in our models, can enhance the CMB heat flow by cooling the mantle and this may be particularly important during Earth's early hot stages \citep{NakagawaTackley2012} and would promote early dynamo operation.
To avoid the formation of an unrealistically large inner core may require a super-heated core or other compensatory heat source.

Several processes are not considered in our models for Earth-sized planets but may drive or modulate dynamos of Solar System bodies of different size and structure.
Mantles much thinner than Earth's allow for efficient core cooling and the growth of an inner core that promotes a chemically-driven dynamo, as has been proposed for present Mercury \citep{Stevenson_etal1983}.
In contrast, a relatively thick mantle and small core may hamper dynamo operation; this has been proposed to explain the absence of a long-lived dynamo on the Moon \citep{Runcorn_etal1975}. 
Although Mars is less massive than Venus, early dynamo cessation due to insufficient CMB heat flow is also a possibility, perhaps triggered by a transition from active tectonics to stagnant lid convection \citep{NimmoStevenson2000}. 
Giant impacts can amplify or generate magnetic fields, which may have contributed to the Moon's paleomagnetic field \citep{HoodVickery1984}.
Mechanical stirring arising from mantle-core differential motion is another possibility for dynamo operation on the early Moon and large asteroids \citep{Dwyer_etal2011}. 
Stable stratification in the outer region of a liquid core, with a dynamo operating only at depth, can buffer a magnetic field and this may cause the weak field of Mercury \citep{Christensen2006}.
A chemically driven dynamo powered through rise or fall of (sulfur-rich) iron snow has been suggested for Ganymede \citep{Bland_etal2008,Hauck_etal2006}.

First-principle computations of liquid iron mixtures \citep{DeKoker_etal2012,Pozzo_etal2012} suggest a thermal conductivity that is 2-3 times higher than previously derived values from extrapolations ($k_c\sim$60 W$^{-1}$ m$^{-1}$, e.g. \citet{StaceyAnderson2001}).
We performed the nominal Earth-like scenario at $k_c=$120 W$^{-1}$ m$^{-1}$ K$^{-1}$.
We demonstrated that in the 60--240 W m$^{-1}$ K$^{-1}$ range, $k_c$ has no influence on the timing of dynamo cessation due to core freeze-out.
However, $k_c\gtrsim$160 W m$^{-1}$ K$^{-1}$ may prevent dynamo operation before inner core formation, which would contradict evidence for a terrestrial magnetic field since at least $\sim$3.5 Ga.
For our stagnant lid scenario SL, lower $k_c$ would likely shorten the magnetic quiet period.
This highlights the importance of 
at least establishing upper limits to the thermal conductivity under core conditions.

%---------------------------------------
\subsection{Implications for exoplanets}

Our results suggest that Earth-mass rocky exoplanets with active surface deformation (plate tectonics or diffuse surface deformation) can have qualitatively similar dynamo evolutions and operate for $\sim$8.2--9.5 billions of years, irrespective of surface temperature (273--759 K).  
A longer-lived ($\sim$10.5 Gyr) dynamo is possible with stagnant lid convection, due to the lower CMB heat flow. 
These dynamo lifetimes raise the question whether the interior dynamics of Earth-mass rocky exoplanets can be inferred by combining magnetic field detections and planet age estimates.
The ages of stars (and their planets) can be inferred with a precision that would be sufficient for identifying planets in different phases of core thermal evolution, i.e. $\sim$1-2 Gyr \citep{MamajekHillenbrand2008}.
Unfortunately, our results suggest that uncertainties in planetary rock properties make a distinction between planets with active surface recycling and stagnant lid planets extremely challenging.
We demonstrated that characterization of exoplanet interior dynamics requires well-determined radius, initial temperature, solidus, and thermal conductivity of the core (Figure \ref{fig-lifetime}).
In particular, the strong sensitivity to the solidus emphasizes the need for constraining the equation of state of iron under core pressures and temperatures.

What dynamo behavior can be expected for planets more massive than Earth?
For rocky planets larger than $\sim$2 Earth masses, it takes $>$4.5 Gyr to form an inner core and dynamo action before that time must be maintained by thermal convection \citep{Gaidos_etal2010,Driscoll2011}.
The pressure-dependence of mantle viscosity is still debated \citep{Stamenkovic_etal2011,Karato2011} but has important implications for mantle dynamics.
Our results demonstrate that a viscosity increase of one order of magnitude across the mantle depth will hamper deep mantle convection and attenuate but not stop dynamo operation. 
Because terrestrial planets more massive than Earth have higher mantle pressures, their material properties probably depart more radically from terrestrial values.
This makes it even more challenging to reliably predict dynamo lifetimes for massive Super-Earths.

%---------------------------------------
\subsection{Detection of magnetic fields on exoplanets?}

Charged particle belts in the magnetospheres of Earth and the giant planets emit
electron cyclotron emission at MHz frequencies.
Magnetodynamo-hosting planets around other stars could presumably emit
at similar frequencies and power levels \citep{Hess2011b}.  Searches
for Jupiter-like emission from giant exoplanets have been undertaken
\citep{Bastian2000,Lazio2007,George2007,Farrell2004,Lazio2010a,Lazio2010b,Lecavelier2011}.
However, predicted frequencies for Earth-like dynamos are below
the 10~MHz cutoff of Earth's ionosphere.  Moreover, an empirical
relation between emitted power and magnetic moment \citep{Zarka2007}
predicts that power emission is well below the detection threshold of
even the most sensitive radio telescope (LOFAR)
\citep{Farrell2004,Driscoll2011,Lazio2010b}.  Detection would require a
fortuitous combination of increased stellar activity
\citep{Griessmeier2005}, beamed emission in the direction of Earth, or
intensification by a stellar flare \citep{Driscoll2011}.  Detection
might also be possible with radio telescopes in space, which would not
suffer from the low-frequency ionospheric cutoff.

The magnetic field of a planet on a close-in orbit can interact with
the field of its host star; periodic chromospheric emission from the
host stars of some ``hot Jupiters'' could be the result of
connectivity between the fields of the planet and the star
\citep{Shkolnik2008,Lanza2011}.  
Other observable phenomenae are associated with a planetary dynamo: 
A magnetic field could alter the circulation and temperature distribution of a partially ionized upper atmosphere of a ``hot'' Earth \citep{Castan2011}, perhaps changing the pattern of reflected or emitted light, although these effects have yet to be explored.  
A magnetized ``tail'' of gas escaping from a
planet could polarize transmitted light \citep{Tachinami2011} or a
bowshock could produce an asymmetry or variability in the 
in the primary occultation signal of a transiting planet
 \citep{Vidotto2011a,Vidotto2011b}

The persistence of an atmosphere on a low-mass planet close to its host star is possible indirect evidence for a planetary dynamo, because a strong magnetic field protects the planet's atmosphere against erosion by a stellar wind and coronal mass ejections.
The dense plasma environment close to the host star can remove hundreds of bars equivalent atmosphere from an unmagnetized Earth-size planet over several Gyr \citep{Khodachenko2007,Lammer2007}.  
Any primary atmosphere dominated by H$_2$ and He would be rapidly removed unless protected by a magnetic field.  
The presence of an atmosphere on a transiting planet can be detected by absorption features in spectra obtained during a transit \citep{Charbonneau_etal2002,Bean2010} or by its re-distribution of heat \citep{Gaidos2004,Knutson2007}. 
Limitations to using planetary atmospheres to infer a magnetodynamo are that planets can conceivably accrete without atmospheres \citep{Raymond2007,Lissauer2007} or may have lost their atmospheres early in their evolution due to increased stellar activity and before any dynamo started.
We calculated an initial non-magnetic period of $\sim$1--1.5 Gyr for all our model scenarios (Table \ref{tab-models}) due to high rates of radiogenic mantle heating, 
which leaves all such planets vulnerable to loss of an atmosphere.
Secondary atmospheres of volcanic CO$_2$, H$_2$O, and N$_2$ may form but could also be removed by thermal escape and stellar wind erosion \citep{Khodachenko2007,Tian2009}.
These possible complications will challenge attempts to infer the operation of a dynamo through the presence or absence of atmospheres on exoplanets.

%%%%%%%%%%%%%%%%%%%%%%%%%%%%%%%%%%%%%%%%
\begin{acknowledgments}
This research was supported by NSF grant EAR-0855546 and NASA grant NNX10AI90G.
We thank Mark Wieczorek, Paul Tackley, and an anonymous reviewer for constructive comments that helped to improve this paper.
\end{acknowledgments}

%%%%%%%%%%%%%%%%%%%%%%%%%%%%%%%%%%%%%%%%

\end{article}

\newpage
%%%%%%%%%%%%%%%%%%%%%%%%%%%%%%%%%%%%%%%%%%%%%%%%%%%%%%%%%%%%%%%%%
\begin{table}
  \caption{
Planetary core parameters and values used in this study. Subscript conventions are ``0'' for zero pressure, ``srfc'' for surface, ``m'' for mantle, ``c'' for core, ``cen'' for planetary center, ``CMB'' for core-mantle boundary, and ``i'' for inner core.  
Values annotated with an asterisk ($^*$) are specific for the nominal-like Earth model C (Table \ref{tab-models}).
Values are from
(a) \citep{Wang_etal2002},
(b) \citep{DeKoker_etal2012,Pozzo_etal2012},
(c) \citep{AndersonAhrens1994},
(d) \citep{DziewonskiAnderson1981},
(e) \citep{PoirierShankland1993},
(f) \citep{DziewonskiAnderson1981,Labrosse2003}, and
(g) \citep{Buffett_etal1996}.
  }
\label{tab-notation}
\begin{tabular}{llll}
\hline
\textbf{Quantity} & \textbf{Symbol} & \textbf{Value} & \textbf{Unit} \\
\hline
$B$            & Magnetic field strength   &                                & \\
$c_{P_c}$      & Core specific heat        & $^*$850$^{(a)}$                & J kg$^{-1}$ K$^{-1}$ \\
$D_{cen} = \sqrt{\frac{3c_P}{2 \pi \alpha \rho_{cen} G}}$
               & Thermal length scale      & $^*$6.17$\times10^{6}$         & - \\
$f=\frac{r_i}{r_{CMB}}$
               & Ratio inner-to-outer core radius
                                           &                                & - \\
$G$            & Gravitational constant    & $6.67\times10^{-11}$
                                                                            & m$^{-3}$ kg$^{-1}$ s$^{-2}$\\
$k_c$          & Heat conductivity core    & 120$^{(b)}$                    & W m$^{-1}$ K$^{-1}$ \\
$K_0$          & Incompressibility at zero pressure
                                           & 110$^{(c)}$                    & GPa \\
$L = \sqrt{ \frac{9 K_0}{2 \pi G \rho_0 \rho_{cen}}\left( \ln (\frac{\rho_{cen}}{\rho_0}) + 1\right)}$
               & Density length scale      & $^*$6.55$\times10^{6}$ & - \\
$M_c$          & Core mass                 & $^*$1.8$\times10^{24}$         & kg \\
$r_i$          & Inner core radius         &                                & m \\
$r_{CMB}$      & CMB radius                & $^*$3.48$\times10^{6}$$^{(d)}$   & m \\
$T_{sol}(r=0)$ & Solidus temperature Fe at center
                                           & 4816                           & K \\
$\bar{T}$      & Effective dissipation temperature & 3500                   & K \\
$\Delta S$     & Entropy of fusion         & 118$^{(e)}$                    & J mol$^{-1}$ K$^{-1}$ \\
$\alpha_{cen}$ & Thermal expansivity at center  & $^*$1.3$\times10^{-5}$$^{(f)}$ & K$^{-1}$ \\
$\bar{\rho}$   & Average core density      &                                & kg m$^{-3}$ \\
$\Delta \rho$  & Density jump at inner-outer core boundary   & 400$^{(g)}$  &  kg m$^{-3}$ \\
$\rho_0$       & Iron density at zero pressure  & 7019$^{(c)}$              & kg m$^{-3}$ \\
$\gamma$       & Gr\"{u}neisen parameter   & $^*$1.2                        & - \\
$\lambda = 1/(\mu_0 \sigma)$
               & Magnetic diffusivity      & 2                              & m$^{2}$ s$^{-1}$ \\
$\mu_0$        & Permeability of free space & 4 $\pi \times 10^{-7}$        & N A$^{-2}$ \\
$\Phi$         & Entropy production rate &  & W K $^{-1}$ \\
               & associated with ohmic dissipation      &         &  \\
\hline
\end{tabular}
\end{table}
%%%%%%%%%%%%%%%%%%%%%%%%%%%%%%%%%%%%%%%%%%%%%%%%%%%%%%%%%%%%%%%%%

%%%%%%%%%%%%%%%%%%%%%%%%%%%%%%%%%%%%%%%%%%%%%%%%%%%%%%%%%%%%%%%%%
\small
\begin{table}
\caption{Terms of the energy ($Q$) and entropy ($E$) balance (Section \ref{Sect-coremodel}).
Subscripts S, L, G, and K refer to secular cooling, latent heat release, gravitational energy release, and conductive cooling, respectively.  Other symbols are given in Table \ref{tab-notation}.  The quantity $F$ in the gravitational terms is $\left( \frac{1}{5} + \frac{2}{15} f^5 - \frac{f^2}{3} \right)  \frac{f}{1-f^3}$, with $f=r_i/r_{CMB}$ the ratio of inner to outer core radius.
Expressions are from (a) \cite{Gaidos_etal2010} and (b) \cite{Nimmo2009} and references therein.
}
\label{tab-terms}
\begin{tabular}{l|l}
\hline
& Energy \\
\hline
$Q_S^{(a)}$
  &
$M_c c_{P_c}
\left[
1
+\frac{2}{5} \left( \frac{r_{CMB}}{D_{cen}} \right)^2
+\frac{4}{35} \left( \frac{r_{CMB}}{D_{cen}} \right)^4
+\frac{12}{175} \left( \frac{r_{CMB}^2}{D_{cen}L} \right)^2
\right]
\frac{dT_{CMB}}{dt}$
 \\
$Q_L^{(a)}$
  &
$\frac{-3M_c \Delta S f}{2(\Delta-1)} \left( \frac{D_{cen}}{r_{CMB}} \right)^2
exp\left(\frac{r_{CMB}^2-r_{ic}^2}{D_{cen}^2}\right)
\frac{dT_{CMB}}{dt}$
  \\
$Q_G^{(b)}$
  &
$\frac{-3\pi G \bar{\rho} M_c F D_{cen}^2}{T_{CMB}}
\frac{\Delta \rho}{\rho (\Delta-1)}
\frac{dT_{CMB}}{dt}$
  \\
$Q_K^{(b)}$
  &
$\frac{6 M_c k_c T_{CMB}}{\bar{\rho} D_{cen}^2}$
  \\
\hline
& Entropy \\
\hline
$E_S^{(a)}$
  &
$\frac{-2M_c c_{P_c}}{5T_{CMB}}
\left(\frac{r_{CMB}}{D_{cen}} \right)^2
\left[
1 + \frac{2}{7} \left( \frac{r_{CMB}}{D_{cen}} \right)^2
+\frac{6}{35} \left( \frac{r_{CMB}}{L} \right)^2
\right]
\frac{dT_{CMB}}{dt}$
\\
$E_L^{(a)}$
  &
$\frac{-3 M_c \Delta S} {2T_{CMB}(\Delta-1)}
\frac{\rho_{ic}}{\bar{\rho}}
f(1-f^2)
\left[ 1+\left(\frac{r_{CMB}}{D_{cen}} \right)^2 \frac{1-f^2}{2} \right]
\frac{dT_{CMB}}{dt}$
  \\
$E_G^{(b)}$
  &
$\frac{Q_G}{T_{CMB}}$
  \\
$E_K^{(b)}$
  &
$\frac{12 M_c k_c r_{CMB}^2}{5 \bar{\rho} D_{cen}^4}\left(1-f^5\right)$
  \\
\hline
\end{tabular}
\end{table}
\normalsize
%%%%%%%%%%%%%%%%%%%%%%%%%%%%%%%%%%%%%%%%%%%%%%%%%%%%%%%%%%%%%%%%%

%%%%%%%%%%%%%%%%%%%%%%%%%%%%%%%%%%%%%%%%%%%%%%%%%%%%%%%%%%%%%%%%%
\begin{table}
\caption{
Specification for the 5 mantle convection models C, H, A, SL, and VI (Section \ref{Sect-cases}). 
Surface temperatures $T_{srfc}$ are uniform across the planet, except for model A, where the surface temperature is 1073 K at the substellar point and decreases sinusoidally toward the terminus and is kept constant on the night-side at 273 K. 
For model SL, pseudo-plastic rheology is excluded ($\sigma_y$ does not apply). 
Model VI differs from model C by having an exponential viscosity increase of a factor 10 across the mantle depth range, in addition to temperature-dependent viscosity.
Magnetodynamo diagnostics are 
onset time of inner core formation ($t_{ons}$), 
time of dynamo cessation due to core freeze-out ($t_{mag}$), 
onse time of thermal stratification ($t_{strat}$, only applies if $<t_{mag}$),
and the maximum intensity of the dipole field at the CMB ($B_{CMB}^{max}$).
}
\label{tab-models}
\begin{tabular}{ll|llllll}
\hline
\textbf{Model Name}  &
\textbf{Abbrev.}  &
\textbf{$T_{srfc}$}  &
\textbf{$\sigma_y$}  &
\textbf{$t_{ons}$}&
\textbf{$t_{mag}$}&
\textbf{$t_{strat}$}&
\textbf{$B_{CMB}^{max}$}\\
\textbf{}  &
\textbf{}  &
\textbf{[K]}  &
\textbf{[MPa]}  &
\textbf{[Gyr]}&
\textbf{[Gyr]}&
\textbf{[Gyr]}&
\textbf{[$\mu$T]}\\
\hline
Cold                       & C   & 273      & 150 & 4.1 & 8.2 & N.A.& 326    \\
Hot                        & H   & 1073     & 150 & 4.4 & 8.7 & 8.4 & 296    \\
Asymmetric                 & A   & 273--1073 & 150 & 4.2 & 8.2 & N.A.& 313    \\
Depth-Increasing Viscosity & VI  & 273      & 150 & 5.0 & 9.3 & 9.0 & 309    \\
Stagnant Lid               & SL  & 273      & N.A.& 5.0 & 10.6 & 4.4 & 252    \\
\hline
\end{tabular}
\end{table}
%%%%%%%%%%%%%%%%%%%%%%%%%%%%%%%%%%%%%%%%%%%%%%%%%%%%%%%%%%%%%%%%%

\newpage
%%%%%%%%%%%%%%%%%%%%%%%%%%%%%%%%%%%%%%%%%%%%%%%%%%%%%%%%%%%%%%%%%
%%%%%%%%%%%%%%%%%%%%%%%%%%%%%%%%%%%%%%%%%%%%%%%%%%%%%%%%%%%%%%%%%

%%%%%%%%%%%%%%%%%%%%%%%%%%%%%%%%%%%%%%%%%%%%%%%%%%%%%%%%%%%%%%%%%
%%%%%%%%%%%%%%%%%%%%%%%%%%%%%%%%%%%%%%%%%%%%%%%%%%%%%%%%%%%%%%%%%%

%%%%%%%%%%%%%%%%%%%%%%%%%%%%%%%%%%%%%%%%%%%%%%%%%%%%%%%%%%%%%%%%%
%%%%%%%%%%%%%%%%%%%%%%%%%%%%%%%%%%%%%%%%%%%%%%%%%%%%%%%%%%%%%%%%%

%%%%%%%%%%%%%%%%%%%%%%%%%%%%%%%%%%%%%%%%%%%%%%%%%%%%%%%%%%%%%%%%%%
%%%%%%%%%%%%%%%%%%%%%%%%%%%%%%%%%%%%%%%%%%%%%%%%%%%%%%%%%%%%%%%%%

%%%%%%%%%%%%%%%%%%%%%%%%%%%%%%%%%%%%%%%%%%%%%%%%%%%%%%%%%%%%%%%%%
%%%%%%%%%%%%%%%%%%%%%%%%%%%%%%%%%%%%%%%%%%%%%%%%%%%%%%%%%%%%%%%%%

%%%%%%%%%%%%%%%%%%%%%%%%%%%%%%%%%%%%%%%%%%%%%%%%%%%%%%%%%%%%%%%%%
%%%%%%%%%%%%%%%%%%%%%%%%%%%%%%%%%%%%%%%%%%%%%%%%%%%%%%%%%%%%%%%%%%

%%%%%%%%%%%%%%%%%%%%%%%%%%%%%%%%%%%%%%%%%%%%%%%%%%%%%%%%%%%%%%%%%%
%%%%%%%%%%%%%%%%%%%%%%%%%%%%%%%%%%%%%%%%%%%%%%%%%%%%%%%%%%%%%%%%%%

\end{document}